\documentclass[aps,prd,showpacs,floatfix,reprint,nofootinbib]{revtex4-1}
\usepackage[dvipdfmx]{graphicx}

\usepackage{bm}
\makeindex

\newcommand \cc[1]{{\bm {#1}}}
\newcommand \cb[1]{\overline{\bm {#1}}}

\begin{document}

\title{Spin-Flavor $\bm{SU(6)}$ Symmetry for Baryon-Baryon Interactions}\label{oka}

\author{Makoto Oka}

\address{Advanced Science Research Center, Japan Atomic Energy Agency, 
Tokai 319-1195, Japan \\
and\\
Nishina Center for Accelerator-Based Science, RIKEN, Wako 351-0198, Japan\\
makoto.oka@riken.jp}

\begin{abstract}
Short-range parts of the baryon-baryon ($BB$) interactions are analyzed 
from the spin-flavor $SU(6)_{sf}\supset SU(3)_f \times SU(2)_s$ symmetry viewpoint.
Due to the Pauli principle of quarks, the symmetry structure of the wave functions is restricted at short distances.
Consequently, the $BB$ states with the same spin-flavor quantum numbers may be reduced into one or a few
spin-flavor states. Such reduction causes repulsion and/or suppression of transitions at short distances.
We show that the observed suppression of the $\Xi N \to \Lambda\Lambda$ conversion
can be explained following the above argument. 
It is also applied to the suppression of the $\Sigma N$ to $\Lambda N$ conversion in the spin 1 and isospin 1/2
channel. 
Furthermore, the effects of the color-magnetic interaction (CMI), which prefers flavor antisymmetric states, 
to the Pauli-allowed states are discussed.
\end{abstract}

\maketitle

\section  {Introduction}
\label{sec:Introduction}

The nuclear force is a fundamental and central subject of nuclear physics and has been intensively studied for several decades.
Recent developments both in experiment and computational physics allow us to access generalized 
nuclear forces, that is, interactions involving baryons with strangeness (hyperons, designated in general by $Y$), 
such as $\Lambda$, $\Sigma$, $\Xi$ and
$\Omega$ baryons. Studies of the generalized nuclear force play important roles in the several aspects as follows. 
(1) Extension to flavor $SU(3)_f$ will help us to reveal the origin and structure of the various components of the nuclear force\cite{Oka:1986fr,generalized-nuclear-force}.
(2) $YN$ and $YY$ interactions are the most basic elements of the dynamics of hyper-nuclei\cite{hypernuclear physics,Hiyama-Nakazawa}.
(3) Strange baryons may play critical roles in dense nuclear matter that can be realized in compact stars in the universe\cite{neutronstar}.

Analyses of the generalized nuclear force are carried out by several different approaches.  
Most traditional approach is to construct a phenomenological potential based 
on the meson exchange mechanism\cite{Julich model,Nijmegen model}. 
There the long-range part of the potential is given by the exchange of the low lying mesons,
while the short-range part is mostly phenomenological. The coupling constants and the short-range
parameters are either fitted to experimental data or determined from the $NN$ parameters by
extrapolation using the $SU(3)_f$ symmetry.

A modern alternative is to formulate the effective field theory\cite{chiral EFT}, where the systematic expansion in terms of
the transferred momentum is introduced. The bound or scattering states of two-baryon systems are 
calculated order by order according to the chiral perturbation theory. 
The parameters of the effective Lagrangian are fitted to the experimental data.

The first-principle calculation by the lattice QCD is also available. 
The HAL-QCD group has established 
a new method to extract the phase-shift equivalent (non-local) potential for two hadron systems
and has applied to general $BB$ interactions\cite{HALQCD}.
This method has an advantage in calculating $BB$ interactions
which are currently not directly accessible by experiment, 
such as multi-strange systems like $N\Omega$, $\Omega\Omega$ or $\Xi\Xi$\cite{exotic-HALQCD}.

The results of these approaches can be compared with experimental data of 
$YN$ scattering experiments\cite{YNexp}, femtoscopy study in heavy ion collisions\cite{Femtoscopy} and 
spectroscopy of hyper-nuclei\cite{hypernuclear physics,Hiyama-Nakazawa}.

The flavor $SU(3)_f$ symmetry is an approximate symmetry of the quantum chromodynamics (QCD). 
In the baryon spectrum, the lowest-lying baryons, $N$, $\Lambda$, $\Sigma$ and $\Xi$, form the  
octet representation of flavor $SU(3)_f$. 
The $SU(3)_f$ is broken by the quark mass term, where the strange quark is heavier than the up and down quarks significantly.
The mass spectrum (a l\'a Gell-Mann and Okubo) and 
the magnetic moments are given consistently with 
the perturbative estimate of the $SU(3)_f$ breaking mass term and the wave functions based on the
$SU(6)_{sf}\supset SU(3)_f\times SU(2)_s$ symmetry.
Because the quark obeys the Fermi-Dirac statistics, the $SU(6)_{sf}$ symmetry is directly connected to 
the symmetry of the color and orbital motion for systems consisting of quarks. The ground state 
baryons belong to the $\cc{56}$ representation, that is the totally symmetric representation designated
by the Young partition number [3]. Note that the color-singlet state of 3 quarks is totally antisymmetric in $SU(3)_c$.

The $SU(6)_{sf}$ symmetry plays an important role also for two-baryon systems.
Now, we consider two-baryon systems consisting of the ground-state baryons belonging to
the [3] symmetric, $\cc {56}$-plet $SU(6)_{sf}$ representation\cite{QCM}.
There the symmetry of 6 quarks will form symmetries given by the Young partition numbers,
\begin{eqnarray}
&& [3]\otimes [3] = [6]_S \oplus [42]_S \oplus [51]_A \oplus [33]_A,
\end{eqnarray}
where $[6]_S$ and $[42]_S$ are symmetric under the exchange of two baryons, while
$[51]_A$ and $[33]_A$ are antisymmetric.
The positive-parity two-baryon states, with $L= 0 \hbox{ or even}$ partial waves, must belong to 
the antisymmetric $[33]_A$ or $[51]_A$.

\begin{widetext}

\begin{table}[htp]
\caption{Possible symmetry combinations of positive-parity and color-singlet 6-quark systems.
}
{\begin{tabular*}{\textwidth}{c@{\extracolsep{\fill}}c@{\extracolsep{\fill}}c} \toprule
Orbital symmetry& Color symmetry & $SU(6)_{sf}$ symmetry\\ \colrule
{}[6] & [222] & [33]\\
{}[42] & [222] & [33]\\
{}[42] & [222] & [51]\\ \botrule
\end{tabular*}}
\label{symtable}
\end{table}%
\end{widetext}

In order to satisfy the Fermi-Dirac statistics, the symmetry combinations are restricted.
Table \ref{symtable} shows possible symmetry combinations for the positive-parity color-singlet states of 6 quarks.
One sees that the color-singlet 6-quark state takes the [222] symmetry, and
the lowest-energy state with the orbital [6] symmetry must have [33] $SU(6)_{sf}$ symmetry.
Such a totally symmetric orbital state will dominate over the mixed symmetric [42] configurations, 
when two baryons come very close
and their wave functions overlap completely (in the relative $L=0$ channel).
Namely, all the quarks reside in the lowest orbit within a single potential well, that is the limit of 
quark shell model\cite{Arima}.
On the other hand, the [51] $SU(6)_{sf}$ symmetry is forbidden for the totally symmetric 
orbital state by the Pauli principle.
This will cause that the [51] state acquires a strong short-range repulsion.
In this way, we can study the short-range interactions of two-baryon systems 
from their $SU(6)_{sf}$ symmetry structure.

When the overlap of two baryons is not complete, 
the orbital mixed symmetry [42] with $SU(6)_{sf}$ [33] and [51] representations will contribute for the relative $L=0$ motion.
The roles of the orbital [42] and $SU(6)_{sf}$ states have been discussed in the context of $NN$ repulsion 
obtained in the quark cluster model\cite{Harvey,Faessler, Arima}.
The relative wave function between the two baryons in the orbital [42] symmetry 
will have a node at the interaction range so that 
it is orthogonal to the totally symmetric [6] wave function.
The kinetic energy becomes larger, such as $2\hbar \omega$ excitation in the harmonic oscillator potential,
and thus an effective repulsion is induced.

In this paper, we discuss the structures of the two-baryon systems in the $SU(6)_{sf}\supset SU(3)_f\times SU(2)_s$
scheme and their significance for the baryon-baryon interactions. 
In Sect.~II, we classify the two-baryon states in the symmetry basis states. 
In order to fully realize the $SU(6)_{sf}$ basis states, we need to include all possible states of two baryons 
consisting with the spin $1/2$ octet ($N$, $\Lambda$, $\Sigma$ and $\Xi$) baryons 
as well as the spin $3/2$ decuplet ($\Delta$, $\Sigma^*$, $\Xi^*$ and $\Omega$) baryons.

In Sect.~III, we discuss the 
effects of the Pauli forbidden states for several two baryon systems.
We discuss how the short-range repulsion is induced by the dominating Pauli-forbidden states. 
We also present the cases where the transition between different two-baryon configurations
at short distances.

In Sect.~IV, we discuss the effect of $SU(6)_{sf}$ breaking, which is important because the different 
$SU(3)_f$ representations within the same $SU(6)_{sf}$ multiplet will split due to the color-spin dependent
interaction of quarks.
We here study the $SU(6)_{sf}$ breaking effects for the two-baryon systems.

Summary and conclusion is given in Sect.~V.

\section{Classification according to $\bm{SU(6)_{sf}\supset SU(3)_f\times SU(2)_s}$}

We consider systems of two $SU(6)_{sf}$ $\cc {56}$-plet baryons based on the constituent quark model.
The quark model hamiltonian is $SU(6)_{sf}$ invariant if we neglect both the quark mass differences
and the spin dependences.

The mass differences of $u$, $d$ and $s$ quarks breaks the $SU(3)_f$ symmetry.
In the quark model, their main contribution is a constant shift of
the hadron masses, which is supplemented by the kinetic energy term proportional to $1/m_q$.
The spin-independent interactions, such as the quark confinement force and the color-electric gluon exchange,
do not depend on flavor.
Thus the effect of the $SU(3)_f$ breaking to the baryon-baryon interaction, 
where the number of strange quarks remains constant, plays a minor role.
So, in the present analysis, we neglect the effect of the $SU(3)_f$ symmetry breaking.

On the other hand, the $SU(6)_{sf}$ symmetry is broken by the spin dependent interaction among the quarks.
It is known that the color-magnetic gluon exchange interaction induces the color-spin dependent interaction, which
gives a splitting between the spin-1/2 octet ($N$, $\Lambda$, $\Sigma$ and $\Xi$) baryons and 
the spin-3/2 decuplet ($\Delta$, $\Sigma^*$, $\Xi^*$, and $\Omega$) baryons. 
We discuss the roles of the color-magnetic interaction in Sect.~IV.


In analyzing the baryon-baryon interaction, we first consider the $SU(6)_{sf}$ and $SU(3)_f$ symmetric limit.
To apply the $SU(6)_{sf}$ bases to two-baryon configurations, we need to 
decompose them into $SU(3)_f$ representations.
The ground state baryons belong to the $SU(6)_{sf}$ [3]-symmetric, 56-dim.\ representation,
${\bf 8}$ ($S=1/2$) and ${\bf 10}$ ($3/2$).
Then the two-baryon systems are classified into the representations given by
\begin{eqnarray}
&& {} {\bf 8} \otimes {\bf 8} = {\bf 1}_S \oplus {\bf 8}_S \oplus {\bf 27}_S \oplus {\bf 8}_A \oplus {\bf 10}_A
\oplus \overline{\bf 10}_A,
\label{88SU3}\\
&& {} {\bf 8} \otimes {\bf 10} = {\bf 8} \oplus {\bf 10} \oplus {\bf 27} \oplus  {\bf 35}, \\
&& {} {\bf 10} \otimes {\bf 10} =  {\bf 27}_S \oplus {\bf 28}_S \oplus \overline{\bf 10}_A \oplus {\bf 35}_A ,
\end{eqnarray}
where the subscript $S$ ($A$) stands for symmetric (antisymmetric) states under the exchange of two baryons.
The $NN$ interactions belong to either $\cb{10}_A$ (for $I=0$) or $\cc{27}_S$ (for $I=1$) representations.
All the other 4 representations, $\cc1$, $\cc 8_S$ and $\cc 8_A$ and $\cc{10}$ in Eq.(\ref{88SU3}) 
are realized only when one or two strange baryons are involved.
In particular, the flavor singlet $\cc1$ representation is formed by the total strangeness $-2$ channels, 
$\Lambda\Lambda$, $\Sigma\Sigma (I=0)$ and $N\Xi (I=0)$.
With the decuplet baryons, we also have new representations, $\cc{28}$ and $\cc{35}$.

\begin{widetext}

\begin{table}[h]
\begin{center}
\caption{Possible $SU(3)_f$ (designated by the dimension of the representation) and spin $S$ states for the $SU(6)_{sf}$ [33] (490-dim.) representation.
See Table \ref{SU3casimirtab} for the dimension ${\bm D}$ of $SU(3)_f$ and
$\Delta=\Delta(S, F)$ defined by Eq.(\ref{MSF}). The corresponding combinations of 
the strangeness ${\cal S}$ and the isospin $I$ are shown in the last column.}
{\begin{tabular}{c|c|r|l}
\hline
${\bm D}$ &$S$ &$\Delta$&$({\cal S},I)$\cr
\hline
$\cc{1}$ & 0 & $-24$ & $(-2,0)$ \cr
$\cc{27}$ & 0 & $8$ & $(0,1), (-1,\frac{1}{2}), (-1,\frac{3}{2}), (-2,0), (-2,1), (-2,2),(-3,\frac{1}{2}),(-3,\frac{3}{2}),(-4, 1)$\cr
$\cc{28}$ & 0 & 48 &$ (0,3), (-1,\frac{5}{2}), (-2,2),(-3,\frac{3}{2}),(-4,1),(-5,\frac{1}{2}), (-6,0)$  \cr
\hline
$\cc{8}$ & 1 & $-28/3$ & $(-1,\frac{1}{2}), (-2,0),(-2,1), (-3,\frac{1}{2})$\cr
$\cc{10}$ & 1 & $8/3 $& $(-1,\frac{3}{2}), (-2,1), (-3,\frac{1}{2}), (-4,0)$ \cr
$\cb{10}$& 1 & $8/3$ & $(0,0), (-1,\frac{1}{2}), (-2,1), (-3,\frac{3}{2})$\cr
$\cc{35}$ & 1 & $80/3$& $ (0,2), (-1,\frac{3}{2}), (-1,\frac{5}{2}),(-2,1),(-2,2),(-3,\frac{1}{2}),(-3,\frac{3}{2}),(-4,0), (-4,1),(-5,\frac{1}{2})$\cr
\hline
$\cc{8}$ & 2 & $4$ &$(-1,\frac{1}{2}), (-2,0),(-2,1),(-3,\frac{1}{2})$ \cr
$\cc{27}$ & 2 & $16$ & $(0,1), (-1,\frac{1}{2}), (-1,\frac{3}{2}), (-2,0), (-2,1), (-2,2),(-3,\frac{1}{2}),(-3,\frac{3}{2}),(-4, 1)$ \cr
\hline
$\cb{10}$& 3 & $16$ & $(0,0), (-1,\frac{1}{2}), (-2,1), (-3,\frac{3}{2})$\cr
\hline
\end{tabular}}
\label{33table}
\end{center}
\end{table}

\begin{table}[h]
\begin{center}
\caption{Possible $SU(3)_f$ and spin states for the $SU(6)_{sf}$ [51] (1050-dim.) representation.}
{\begin{tabular}{c|c|r|l}
\hline
${\bm D}$ &  $S$ &$\Delta$&$({\cal S},I)$\cr
\hline
$\cc{8}$ & 0 & $-12$& $(-1,\frac{1}{2}), (-2,0),(-2,1), (-3,\frac{1}{2})$\cr
$\cc{27}$ & 0 & $8$&$(0,1), (-1,\frac{1}{2}), (-1,\frac{3}{2}), (-2,0), (-2,1), (-2,2),(-3,\frac{1}{2}),(-3,\frac{3}{2}),(-4, 1)$\cr
\hline
$\cc{8}$ & 1 & $-28/3$ & $(-1,\frac{1}{2}), (-2,0),(-2,1), (-3,\frac{1}{2})$\cr
$\cc{10}$ & 1 &$8/3$&$(-1,\frac{3}{2}), (-2,1), (-3,\frac{1}{2}), (-4,0)$\cr
$\cb{10}$& 1 &  $8/3$&$(0,0), (-1,\frac{1}{2}), (-2,1), (-3,\frac{3}{2})$\cr
$\cc{27}$ & 1 & $32/3$&$(0,1), (-1,\frac{1}{2}), (-1,\frac{3}{2}), (-2,0), (-2,1), (-2,2),(-3,\frac{1}{2}),(-3,\frac{3}{2}),(-4, 1)$\cr
$\cc{35}$ & 1 & $80/3$ & $ (0,2), (-1,\frac{3}{2}), (-1,\frac{5}{2}),(-2,1),(-2,2),(-3,\frac{1}{2}),(-3,\frac{3}{2}),(-4,0), (-4,1),(-5,\frac{1}{2})$\cr
\hline
$\cc{10}$ & 2 & $8$&$(-1,\frac{3}{2}), (-2,1), (-3,\frac{1}{2}), (-4,0)$\cr
$\cc{27}$ & 2 & $16$&$(0,1), (-1,\frac{1}{2}), (-1,\frac{3}{2}), (-2,0), (-2,1), (-2,2),(-3,\frac{1}{2}),(-3,\frac{3}{2}),(-4, 1)$\cr
$\cc{35}$ & 2& $32$& $ (0,2), (-1,\frac{3}{2}), (-1,\frac{5}{2}),(-2,1),(-2,2),(-3,\frac{1}{2}),(-3,\frac{3}{2}),(-4,0), (-4,1),(-5,\frac{1}{2})$\cr
$\cc{28}$ & 2& $56$ &$ (0,3), (-1,\frac{5}{2}), (-2,2),(-3,\frac{3}{2}),(-4,1),(-5,\frac{1}{2}), (-6,0)$ \cr
\hline
$\cc{35}$ & 3& $40$& $ (0,2), (-1,\frac{3}{2}), (-1,\frac{5}{2}),(-2,1),(-2,2),(-3,\frac{1}{2}),(-3,\frac{3}{2}),(-4,0), (-4,1),(-5,\frac{1}{2})$\cr
\hline
\end{tabular}}
\label{51table}
\end{center}
\end{table}%
%
\end{widetext}

In Tables \ref{33table} and \ref{51table}, we tabulate the possible two baryon states of $SU(6)_{sf}$ [33] and [51] symmetric representations, respectively.
There one sees some quantum numbers appear only in the [51] $SU(6)_{sf}$ representation. 
For such quantum numbers,
the totally symmetric [6] orbital state is forbidden and therefore we expect a strong short-range repulsion induced by
the Pauli principle.
Such channels include $BB'(S,I)= N\Delta (1,1) $, $ N\Delta (2,2) $, 
$\Delta\Delta(2,3) $, $\Delta\Delta(3,2) $, $(\Delta\Sigma+\Delta\Sigma^* )(2,\frac{5}{2}) $, 
$\Delta\Sigma^* (3, \frac{3}{2}) $,  $\Delta\Sigma^*(3, \frac{5}{2}) $, and so on.

On the other hand, many combinations of quantum numbers, particularly with lower spins, appear both 
in the [33] and [51] $SU(6)_{sf}$ representations.
There a part of the two-baryon channels is forbidden at short distances and 
we need to decompose the two-baryon states into the $SU(6)_{sf} \supset SU(3)_f$ representations. 
We will see that the spin-dependent interaction that breaks the $SU(6)_{sf}$ symmetry becomes important
and then the lower symmetric $SU(3)_f$ representations happen to be dominant.
(In fact, the states with the lower values of $\Delta(S,F)$ in Table \ref{33table} are favored.)
Thus the flavor singlet $\cc{1}$ and octet $\cc{8}$ states will have a good chance to make a compact dibaryon state.

In the following, we take four cases and analyze them from the $SU(6)_{sf} \supset SU(3)_f$ symmetry viewpoint. (1) $BB$ with strangeness ${\cal S}= -2$, spin $S=0$, and isospin $I=0$, (2) $BB$ with ${\cal S}=-1$, $S=0$, $I=1/2$, (3) $BB$ with ${\cal S}=-1$, $S=1$, $I=1/2$ and (4) non-strange $BB$ with $S=0$, $I=1$ and $S=1$, $I=0$. 

\subsection{Strangeness $\bm{{\cal S}=-2}$, spin $\bm{S=0}$, isospin $\bm{I=0}$ states}\label{S=-2}

First we consider the states with strangeness ${\cal S}=-2$, spin $S=0$, and isospin $I=0$.
In the particle basis, they consist of
$\Lambda\Lambda$, $(N \Xi)_S$, $\Sigma\Sigma$, and $\Sigma^*\Sigma^*$.
This system has attracted much attention because of  possible existence of the dibaryon, a compact 6-quark
state predicted by Jaffe in 1970\cite{Jaffe}.
The H dibaryon has attracted much attention, which are searched by experiment\cite{H-search} and also is studied by lattice QCD\cite{H-dibaryon-HALQCD}. 
The quark cluster model was applied to the H dibaryon system\cite{OSYforH}.

For the $SU(3)_f$ basis, they belong to the symmetric $BB$ representations out of
 ${\bf 8}\times{\bf 8}$, and ${\bf 10}\times{\bf 10}$, given in terms of the particle bases as
\begin{eqnarray}
&& |{\bf 1}\rangle = -\sqrt{\frac{1}{8}} |\Lambda\Lambda\rangle + \sqrt{\frac{1}{2}} |N\Xi \rangle_S 
+ \sqrt{\frac{3}{8}} |\Sigma\Sigma\rangle
\\
\nonumber\\
&&  |{\bf 8}\rangle = -\sqrt{\frac{1}{5}} |\Lambda\Lambda\rangle + \sqrt{\frac{1}{5}} |N\Xi \rangle_S 
- \sqrt{\frac{3}{5}} |\Sigma\Sigma\rangle
\\
\nonumber\\
&&  |{\bf 27}\rangle_1 = \sqrt{\frac{27}{40}} |\Lambda\Lambda\rangle + \sqrt{\frac{12}{40}} |N\Xi \rangle_S 
- \sqrt{\frac{1}{40}} |\Sigma\Sigma\rangle
\\
\nonumber\\
&& |{\bf 27}\rangle_2 =  |\Sigma^*\Sigma^*\rangle
\end{eqnarray}

\begin{widetext}

In the $SU(6)_{sf}\supset SU(3)_f \times SU(2)_s$ basis, they are classified into 
the four states,
($[33]\, {\bf 1}$,  $[33]\, {\bf 27}$, $[51]\,{\bf 8}$,  $[51]\,{\bf 27}$),
where the Young partition $[f_1f_2]$ represents the $SU(6)_{sf}$ representation and 
the bold number is for the $SU(3)_f$ multiplet.
The relation between the $SU(6)_{sf}$ basis and the particle basis is given by
\begin{eqnarray}
&& \begin{pmatrix}{
 [33]\, {\bf 1} \cr  [33]\, {\bf 27}\cr [51]\,{\bf 8} \cr[51]\,{\bf 27}\cr
}\end{pmatrix}
=\frac{1}{\sqrt{360}}
\begin{pmatrix}
{
-\sqrt{45}& \sqrt{180}&\sqrt{135} & 0\cr
\sqrt{135}& \sqrt{60}&-\sqrt{5}& -\sqrt{160} \cr
-\sqrt{72}&\sqrt{72} &-\sqrt{216}& 0 \cr
-\sqrt{108}& -\sqrt{48}&\sqrt{4}& -\sqrt{200}\cr
}
\end{pmatrix}
 \begin{pmatrix}
{
\Lambda\Lambda \cr N\Xi_S \cr \Sigma\Sigma \cr \Sigma^*\Sigma^* \cr 
}
\end{pmatrix},
\end{eqnarray}
or conversely, 
\begin{eqnarray}
&&
 \begin{pmatrix}
{
\Lambda\Lambda \cr N\Xi_S \cr \Sigma\Sigma \cr \Sigma^*\Sigma^* \cr 
}
\end{pmatrix}
=\frac{1}{\sqrt{360}}
\begin{pmatrix}
{
-\sqrt{45}&\sqrt{135}& -\sqrt{72} & - \sqrt{108}\cr
\sqrt{180}& \sqrt{60}& \sqrt{72}& -\sqrt{48}\cr
\sqrt{135}&-\sqrt{5} & -\sqrt{216}& \sqrt{4}\cr
0 &-\sqrt{160}& 0 & -\sqrt{200}\cr
}
\end{pmatrix}
 \begin{pmatrix}
{
[33]\, {\bf 1} \cr  [33]\, {\bf 27}\cr [51]\,{\bf 8} \cr[51]\,{\bf 27}\cr 
}
\end{pmatrix}.
\label{-2,0,0}
\end{eqnarray}

\end{widetext}

These classifications can be obtained as the eigenstates of the quark exchange operators.
More concretely, the above expression can be obtained by diagonalizing the matrix elements of the $P_{36}^{sf}$ operator, that
exchanges a quark from the first baryon and a quark from the second. In Appendix I, the $P_{36}^{sf}$ matrices 
for the particle basis states are given explicitly.

\subsection{$\bm{{\cal S}=-1}$, $\bm{S=0}$, $\bm{I=1/2}$ states}\label{S-1,S=0}

Next, we consider the strangeness $-1$ two baryon systems with $S=0$ and $I=1/2$.
They consist of the symmetric combinations, ($(N\Lambda)_S$, $(N\Sigma)_S$, $(\Delta\Sigma^*)_S$),
which are rearranged into the $SU(3)_f$ basis states 
as
\begin{eqnarray}
&& |{\bf 8}\rangle = -\sqrt{\frac{1}{10}} |N\Lambda\rangle_S +  \sqrt{\frac{9}{10}} |N\Sigma\rangle_S
\\
\nonumber\\
&&  |{\bf 27}\rangle_1 = \sqrt{\frac{9}{10}} |N\Lambda\rangle_S +  \sqrt{\frac{1}{10}} |N\Sigma\rangle_S
\\
\nonumber\\
&& |{\bf 27}\rangle_2 =  |\Delta\Sigma^*\rangle_S
\end{eqnarray}
Note that the $\cc8\otimes\cc{10}$ combinations are not allowed for $S=0$ states.

\begin{widetext}
Now similarly to the strangeness $-2$ system, we relate them to the
$SU(6)_{sf}$ basis states, giving
\begin{eqnarray}
&& \begin{pmatrix}
{
[33]\, {\bf 27}\cr [51]\,{\bf 27} \cr  [51]\,{\bf 8} \cr
}
\end{pmatrix}
=\frac{1}{\sqrt{90}}
\begin{pmatrix}
{
\sqrt{45}&\sqrt{5}& -\sqrt{40}\cr
\sqrt{36}&\sqrt{4}& \sqrt{50}\cr
-\sqrt{9}&\sqrt{81}& 0\cr
}
\end{pmatrix}
 \begin{pmatrix}
{
N\Lambda_S \cr N\Sigma_S \cr \Delta\Sigma^*_S \cr
}
\end{pmatrix},
\end{eqnarray}
and conversely, 
\begin{eqnarray}
&& \begin{pmatrix}
{
N\Lambda_S \cr N\Sigma_S \cr \Delta\Sigma^*_S \cr
}
\end{pmatrix}
=\frac{1}{\sqrt{90}}
\begin{pmatrix}
{
\sqrt{45}&\sqrt{36}& -\sqrt{9}\cr
\sqrt{5}&\sqrt{4}& \sqrt{81}\cr
-\sqrt{40}&\sqrt{50}& 0\cr
}
\end{pmatrix}
 \begin{pmatrix}
{
  [33]\, {\bf 27}\cr [51]\,{\bf 27} \cr [51]\,{\bf 8} \cr
}
\end{pmatrix}.
\label{-1,0,1/2}
\end{eqnarray}

\end{widetext}

\subsection{$\bm{{\cal S}=-1}$, $\bm{S=1}$, $\bm{I=1/2}$ states}\label{Strange,S=1}
Now we consider the strangeness $-1$, $S=1$, $I=1/2$ states,
($(N\Lambda)_A$, $(N\Sigma)_A$, $(N\Sigma^*)_A$, $(\Delta\Sigma)_A$, $(\Delta\Sigma^*)_A$).
The $SU(3)_f$ basis states from  ${\bf 8}\times{\bf 8}$, ${\bf 8}\times{\bf 10}$, and ${\bf 10}\times{\bf 10}$ combinations 
are given by
\begin{eqnarray}
&& |{\bf 8}\rangle_1 = \frac{1}{\sqrt{2}} |N\Lambda\rangle_A +  \frac{1}{\sqrt{2}} |N\Sigma\rangle_A
\\
&&  |\overline{\bf 10}\rangle_1 = \frac{1}{\sqrt{2}} |N\Lambda\rangle_A -  \frac{1}{\sqrt{2}} |N\Sigma\rangle_A
\\
&&
|{\bf 8}\rangle_2 = \frac{1}{\sqrt{5}} |N\Sigma^*\rangle_A +  \frac{4}{\sqrt{5}} |\Delta\Sigma\rangle_A
\\
&&
|{\bf 27}\rangle =  \frac{4}{\sqrt{5}} |N\Sigma^*\rangle_A -  \frac{1}{\sqrt{5}} |\Delta\Sigma\rangle_A
\\
&& |\overline{\bf 10}\rangle_2 = |\Delta\Sigma^*\rangle_A. 
\end{eqnarray}

\begin{widetext}
Again the correspondent $SU(6)_{sf}$ basis states are given by
\begin{eqnarray}
&& \begin{pmatrix}
{
[33]\, {\bf 8}\cr [33]\, \overline{\bf 10}\cr [51]\,{\bf 8} \cr [51]\,\overline{\bf 10} \cr  [51]\,{\bf 27} \cr  
}
\end{pmatrix}
=\frac{1}{\sqrt{90}}
\begin{pmatrix}
{
\sqrt{20}&\sqrt{20}& -\sqrt{10}& -\sqrt{40}&0\cr
\sqrt{25}&-\sqrt{25}& 0 & 0 & -\sqrt{40}\cr
\sqrt{25}& \sqrt{25}& \sqrt{8}&\sqrt{32}& 0\cr
\sqrt{20}&-\sqrt{20}& 0 & 0 & \sqrt{50}\cr
0& 0& \sqrt{72}&-\sqrt{18}& 0\cr
}
\end{pmatrix}
 \begin{pmatrix}
{
N\Lambda_A \cr N\Sigma_A \cr N\Sigma^*_A \cr \Delta\Sigma_A \cr \Delta\Sigma^*_A \cr
}
\end{pmatrix},
\end{eqnarray}
and
\begin{eqnarray}
&& \begin{pmatrix}
{
N\Lambda_A \cr N\Sigma_A \cr N\Sigma^*_A \cr \Delta\Sigma_A \cr \Delta\Sigma^*_A\cr
}
\end{pmatrix}
=\frac{1}{\sqrt{90}}
\begin{pmatrix}
{
\sqrt{20}&\sqrt{25}&\sqrt{25} &\sqrt{20} &0\cr
\sqrt{20}&-\sqrt{25}& \sqrt{25} & -\sqrt{20} &0 \cr
-\sqrt{10}&0& \sqrt{8}&0&  \sqrt{72}\cr
-\sqrt{40}&0& \sqrt{32} & 0 &-\sqrt{18} \cr
0& -\sqrt{40}&0&\sqrt{50}& 0\cr
}
\end{pmatrix}
\begin{pmatrix}
{
[33]\, {\bf 8}\cr [33]\, \overline{\bf 10}\cr [51]\,{\bf 8} \cr [51]\,\overline{\bf 10} \cr  [51]\,{\bf 27} \cr  
}
\end{pmatrix}.
\label{eqS=-1}
\end{eqnarray}
\end{widetext}

\subsection{Non-strange $\bm{S=0}$, $\bm{I=1}$  and $\bm{S=1}$, $\bm{I=0}$ states}\label{NS}

Finally, we give the classifications of the $NN$ and $\Delta\Delta$ states with $(S, I)=(0,1)$ and $(1,0)$.
The $(S,I)=(0,1)$ $NN-\Delta\Delta$ states belong to the flavor $\cc{27}$ representation and can be
classified by the $SU(6)_{sf}$ symmetry as
\begin{eqnarray}
&& \begin{pmatrix}
{
NN_S \cr \Delta\Delta_S \cr 
}
\end{pmatrix}
=\frac{1}{3}
\begin{pmatrix}
{
\sqrt{5}&\sqrt{4}\cr
-\sqrt{4}&\sqrt{5}\cr
}
\end{pmatrix}
 \begin{pmatrix}
{
  [33]\, {\cc{27}}\cr [51]\,{\cc{27}}  \cr
}
\end{pmatrix}.
\end{eqnarray}

The $(S,I)=(1,0)$ $NN-\Delta\Delta$ states belong to the flavor $\cb{10}$ representation and are given by
\begin{eqnarray}
&& \begin{pmatrix}
{
NN_A \cr \Delta\Delta_A \cr 
}
\end{pmatrix}
=\frac{1}{3}
\begin{pmatrix}
{
\sqrt{5}&\sqrt{4}\cr
-\sqrt{4}&\sqrt{5}\cr
}
\end{pmatrix}
 \begin{pmatrix}
{
  [33]\, {\cb{10}}\cr [51]\,{\cb{10}}  \cr
}
\end{pmatrix}.
\end{eqnarray}

\section{Short-range interactions of baryons}

Now it is straightforward to see the short-range behavior of the baryon-baryon interaction in the $SU(6)_{sf}$ symmetry limit.
We remember that the Fermi-Dirac statistics allows only the [33] symmetric states when two baryons sit on top of each other.
Thus at the very short distance, the relative wave function of two baryons should be dominated by the [33] component.

As an example, in the case of the $NN-\Delta\Delta (I=1) $ system (see Sect. \ref{NS}), there is only one [33] state, $[33] \cc {27}$.
The [51] component, $[51] \cc{27}$, is forbidden by the Pauli principle and thus will disappear at short distances.
In other words, the $NN$ and $\Delta\Delta$ components, which are independent in the asymptotic region, are reduced into a single 
component and are not distinguishable at short distances. 
There 55\% of the $NN$ wave function is $[33] \cc {27}$, but the remaining 45\% is forbidden by the Pauli principle.

It is natural to ask whether this effect of the Pauli principle explains the short-range repulsion between two nucleons
that was established experimentally and was also demonstrated by the quark cluster model calculations. 
It, however, is found that the 45\% suppression of the $NN$ short-range wave function by the Pauli principle is not enough to induce the short-range repulsion.
To understand the short-range repulsion, we need the $SU(6)_{sf}$ breaking interaction, 
which splits the $NN$ and $\Delta\Delta$, 
as is discussed in the next section.

On the other hand, the effect of antisymmetrization becomes prominent in the ${\cal S}=-1$, $(S, I)=(0, 1/2)$ channel (Sect. \ref{S-1,S=0}), 
where only one [33] state, $[33]\cc{27}$ is allowed at short distances. 
From Eq.(\ref{-1,0,1/2}), one sees that the probability of the allowed state is 50\%, 5.6\% and 10\% for $N\Lambda$, $N\Sigma$ and $\Delta\Sigma^*$, respectively. Therefore the $N\Sigma$ wave function is strongly suppressed (almost forbidden) at short distances. This is a ``quasi''-Pauli-forbidden state, which effectively induces a strong repulsion between the baryons.
Thus it is expected that the $N\Sigma$ interaction for the S-wave spin $S=1/2$ and isospin $I=0$ channel is strongly repulsive.

A similar quasi-Pauli-forbidden state appears also in the $N\Sigma$ state with ${\cal S}=-1$, $(S,I)=(1, 3/2)$ channel.
There are five $S$-wave states, $N\Sigma$, $N\Sigma^*$, $\Delta\Lambda$, $\Delta\Sigma$ and $\Delta\Sigma^*$, and
they form $[33]\cc{10}$, $[33]\cc{35}$, $[51]\cc{10}$, $[55]\cc{27}$ and $[55]\cc{35}$ representations.
The only state with two octet baryons, $N\Sigma$ contains only 11\% of [33] state,
\begin{eqnarray}
&& |N\Sigma (1,3/2)\rangle = \sqrt{\frac{1}{9}} |[33] \cc{10}\rangle  + \hbox{$|[51]$ states$\rangle$ \ldots}.
\end{eqnarray}
Therefore the $S$-wave interaction in the $(S,I)=(1,3/2)$ $N\Sigma$ state is also strongly repulsive.
In contrast, the $N\Sigma$ $(S,I)=(0,3/2)$ channel has only two S-wave states, $N\Sigma$ and $\Delta\Sigma^*$,
which form $[33]\cc{27}$ and $[51]\cc{27}$ representations. The lowest $N\Sigma$ state is decomposed as
\begin{eqnarray}
&& |N\Sigma(0,3/2)\rangle = \sqrt{\frac{5}{9}} |[33] \cc{27}\rangle  + \sqrt{\frac{4}{9}} |[51] \cc{27}\rangle,
\end{eqnarray}
and the Pauli principle in this case does not induce repulsion.

The short-range repulsions between $N$ and $\Sigma$ in the $(S, I)=(0, 1/2)$ and $(1,3/2)$ channels were
demonstrated by the quark
cluster model calculation\cite{Oka:1986fr}.
The strengths of the repulsions are much larger than the repulsion between two nucleons,
showing that the effect  of the Pauli forbidden state is dominant compared with the $SU(6)_{sf}$ breaking effect.
The repulsions in these channels were confirmed also in the lattice computation of the potential\cite{Inoue:2018axd,Nemura:2018tay}.
The E40 experiment at J-PARC was very successful in measuring the $p\Sigma^{\pm}$ scattering cross sections recently\cite{YNexp,J-PARCE40:2022nvq}.
The data are analyzed to extract the S-wave phase shifts, which show moderate $N\Sigma$ repulsion at low and medium energies.
The repulsion of $\Sigma-N$ interaction is also consistent with the experimental observation of the $\Sigma$-nucleus potential\cite{Sigma-Nucleus,Inoue:2018axd}.


\bigbreak
When there are more than one [33] states available, the $SU(6)_{sf}$ classification will give us non-trivial relations.
We consider the strangeness ${\cal S}=-2$, $(S, I)=(0,0)$ case (Sect. \ref{S=-2}). Eq.(\ref{-2,0,0}) shows us that the wave functions at short
distances are dominated by the [33] states, as
\begin{eqnarray}
&& |\Lambda\Lambda\rangle = -\sqrt{\frac{1}{8}} |[33] \cc{1}\rangle + \sqrt{\frac{3}{8}} |[33] \cc{27}\rangle + \hbox{[51] \ldots}\\
&& |N\Xi\rangle_S =\sqrt{\frac{1}{2}} |[33] \cc{1}\rangle +\sqrt{\frac{1}{6}} |[33] \cc{27}\rangle + \hbox{[51] \ldots}\\
&&  |\Sigma\Sigma\rangle = \sqrt{\frac{3}{8}} |[33] \cc{1}\rangle - \sqrt{\frac{1}{72}}  |[33] \cc{27}\rangle + \hbox{[51] \ldots}\\
&&  |\Sigma^*\Sigma^*\rangle =  - \frac{2}{3}  |[33] \cc{27}\rangle + \hbox{[51] \ldots}
\end{eqnarray}
One sees that the probabilities of the [33] states are 50\%, 67\%, 39\% and 44\% for $\Lambda\Lambda$, $N\Xi$, $\Sigma\Sigma$
and $\Sigma^*\Sigma^*$, respectively, and no significant suppression is seen for these channels.
On the other hand, we see that the overlap between $|\Lambda\Lambda\rangle$ and $|N\Xi_S\rangle$ states is zero.
Namely, the [33] components of $|\Lambda\Lambda\rangle$ and $|N\Xi_S\rangle$ are orthogonal to each other.
Within the approximation of $SU(6)_{sf}$ invariance of the inter-quark interaction, 
there is no mixing between $\Lambda\Lambda$ and $N\Xi$ at short distances in this quantum number.
We here conclude that the $N\Xi$ to $\Lambda\Lambda$ conversion is strongly suppressed at
the distances where the quark wave functions of two baryons overlap significantly. 
It explains why the transition potential from $N\Xi$ to $\Lambda\Lambda$ is suppressed at short distances. 
This property of the $N\Xi$-$\Lambda\Lambda$ system has been pointed out by the lattice
QCD calculation\cite{XiN-LL conversion}. 
It is also consistent with the discoveries of $\Xi$-hypernuclei in the J-PARC E07
experiment\cite{Xi-exp,Hiyama-Nakazawa}.

Another interesting case is ${\cal S}=-1$, $(S, I)=(1,1/2)$ (Sect. \ref{Strange,S=1}) system.
One sees two allowed [33] states, $[33]\cc8$ and $[33]\cb{10}$.
The two baryon systems contain these states as
\begin{eqnarray}
&& |N\Lambda\rangle_A = \sqrt{\frac{4}{18}} |[33] \cc{8}\rangle + \sqrt{\frac{5}{18}} |[33] \cb{10}\rangle + \hbox{[51] \ldots}\\
&&\nonumber\\
&& |N\Sigma\rangle_A = \sqrt{\frac{4}{18}} |[33] \cc{8}\rangle - \sqrt{\frac{5}{18}} |[33] \cb{10}\rangle + \hbox{[51] \ldots}\\
&&  |N\Sigma^*\rangle_A = -\sqrt{\frac{1}{9}} |[33] \cc{8}\rangle  + \hbox{[51] \ldots}\\
&&  |\Delta\Sigma\rangle_A =  -\sqrt{\frac{4}{9}} |[33] \cc{8}\rangle  + \hbox{[51] \ldots}\\
&&  |\Delta\Sigma^*\rangle_A =- \sqrt{\frac{4}{9}} |[33] \cb{10}\rangle+ \hbox{[51] \ldots}
\end{eqnarray}
Now, one sees that the $N\Sigma^*$ state is almost 90\% suppressed by the Pauli principle 
so that a strong repulsion is expected there.
We also see that the [33] part of $N\Lambda$ and $N\Sigma$ are nearly orthogonal. 
Again, the $SU(6)_{sf}$ invariance will
suppress the transition between these states at short distances, which again is consistent with
the lattice QCD calculation of the transition potential\cite{Nemura:2018tay}.

\section{Color Magnetic Interaction}

So far, we have classified the two-baryon systems in terms of the $SU(6)_{sf}$ symmetry.
However, this symmetry is broken as the splittings 
between the octet ($N$, $\Lambda$, \ldots) and decuplet ($\Delta$, $\Sigma^*$, \ldots) baryons indicate.
For a compact state of quarks, the most important dynamics that breaks $SU(6)_{sf}$ invariance 
is the spin-color dependent force, 
called color magnetic interaction(CMI). Its origin may be a gluon exchange between constituent quarks\cite{DGG}, or the 
instanton induced interaction for light quarks\cite{OT}.
The CMI is given by 
\begin{eqnarray}
&& V_{\rm CMI} =  - \sum_{i<j} ({\bm \sigma}_i\cdot{\bm \sigma}_j) (\lambda_{i}\cdot\lambda_{j}) v_0(r_{ij})
\end{eqnarray}
where $r_{ij}$ is the distance between the $i$-th and $j$-th quarks, $\lambda_{i(j)}$ denotes the color Gell-Mann matrix for the $i(j)$-th quark.
Here we neglect the mass difference between $u$, $d$ and $s$ quarks, namely considering the $SU(3)_f$ limit.
Assuming for the compact 6-quark state that all the quarks are in the same orbital state, the expectation value is given by
\begin{eqnarray}
&& \langle V_{\rm CMI}\rangle =  \langle v_0 \rangle \Delta(S, F)
\label{CMIeq}
\end{eqnarray}
where $\Delta(S,F)$ depends on the total spin $S$ and the flavor representation, $F=(p,q)$. 
It is given by the formula%
\footnote{
A general formula for fully antisymmetric $n$ quarks with spin $S$, flavor $F$ and color $C$ is given by
\begin{eqnarray}
&& \Delta(S,F,C) = -\frac{2}{3}\frac{n(n-1)}{2} +\frac{4}{3} \langle \sum_{i<j}P_{ij}^{s}\rangle
+ 2\langle \sum_{i<j}P_{ij}^{c}\rangle +4\langle \sum_{i<j}P_{ij}^{f}\rangle\nonumber\\
&& = n(n-10) +\frac{4}{3}S(S+1) +  2C_2[C] + 4C_2[F], \nonumber
\end{eqnarray}
where $P_{ij}^{s,f,c}$ are the exchange operators for the spin ($s$), flavor ($f$) and color ($c$) part of the wave function, respectively. $C_2[C]$ vanishes for color-singlet states.}%
\begin{eqnarray}
&& \Delta(S, F=(p,q)) = \langle - \sum_{i<j} ({\bm \sigma}_i\cdot{\bm \sigma}_j) (\lambda_{i}\cdot\lambda_{j})\rangle\nonumber\\
&& = n(n-10) +\frac{4}{3}S(S+1) +  4C_2[(p,q)],
\label{MSF}
\end{eqnarray}
where $n$ is the total number of quarks ($n=6$ for two baryons), 
$C_2$ is the $SU(3)$ quadratic Casimir invariant of the flavor representation $F=(p,q)$ given in Table \ref{SU3casimirtab}.
The values of $\Delta(S,F)$ are tabulated for the relevant spin and flavor in Tables \ref{33table} and \ref{51table}.%

\begin{widetext}

\begin{table}[h]
\caption{Casimir eigenvalues of the $SU(3)$ representations. 
$[f]=[f_1,f_2,f_3]$ denotes the Young partition numbers for the symmetry
of the representation, which specifies $p=f_1-f_2$ and $q=f_2-f_3$.
Then $p$ and $q$ give the  
dimension ${\bm D}= (p+1)(q+1)(p+q+2)/2$ and the Casimir eigenvalue
$C_2= (p^2+pq+q^2)/3+p+q$ of the representation.
}
\begin{center}
{\begin{tabular*}{\textwidth}{c@{\extracolsep{\fill}} c @{\extracolsep{\fill}}c @{\extracolsep{\fill}}c}
\hline
${\bm D(p,q)}$&$[f]$ &$(p,q)$ &$C_2(p,q)$\\
\hline
$\cc{1}$&$[111]$ &$(0,0)$ &$0$\\ 
$\cc{8}$&$[21]$ &$(1,1)$ &$3$\\
$\cc{10}$&$[3]$ &$(3,0)$ & $6$\\
$\cb{10}$& $[33]$ &$(0,3)$ & $6$\\
$\cc{27}$&$[42]$ &$(2,2)$ &$8$\\
$\cc{35}$&$[51]$ &$(4,1)$ &$12$\\
$\cc{28}$& $[6]$ &$(6,0)$ & $18$\\
 \hline
\end{tabular*}}
\end{center}
\label{SU3casimirtab}
\end{table}%

\end{widetext}

The strength factor, $\langle v_0 \rangle$ in Eq.(\ref{CMIeq}), is positive and thus the smaller 
$SU(3)$ Casimir eigenvalue is preferred. 
As the splitting between the octet and decuplet baryons, $N$ and $\Delta$, is given by $16 \langle v_0\rangle$,
we can make a rough estimate, $\langle v_0\rangle\sim 300/16$ MeV.
The flavor symmetric states have larger Casimir eigenvalues than flavor antisymmetric states, 
the dibaryon states with antisymmetric flavor eigenstates are favored by CMI.
Among the [33] symmetric states, the most favored state by CMI is the $\cc{1} (S=0)$ state, which appears
Eq.(\ref{-2,0,0}). 
According to Table I, $\Delta(S,F)$ is $-24$.
This is the state relevant for the possible H dibaryon, $uuddss$ system with strangeness ${\cal S}=-2$, $(S, I) =(0,0)$.
The other [33] state in Eq.(\ref{-2,0,0}) is $\cc{27} (S=0)$, for which $\Delta(S,F)$ is $+8$, {\it i.e.}, $\sim 600$ MeV above
the $\cc{1} (S=0)$ state. 
Thus we conclude that the compact low-energy state is dominated by the $\cc{1} (S=0)$ state.

Similar analysis can be done for the ${\cal S}=-1$, $(S, I)=(1,1/2)$ states, Eq.(\ref{eqS=-1}).
The two [33] states, $\cc8 (S=1)$ and $\cb{10}(S=1)$, have $\Delta(S,F)=-28/3$ and $8/3$, and thus
the $\cc8 (S=1)$ is about 225 MeV above the $\cb{10}(S=1)$.
Both the $N\Lambda$ and $N\Sigma$ states contain the $\cc8 (S=1)$ and  $\cb{10}(S=1)$ states by
22 and 28\%, respectively, though the relative signs are opposite.
So, here one of the combinations of $N\Lambda$ and $N\Sigma$ becomes dominant at short distances, 
while the overall mixing of the $N\Lambda$ and $N\Sigma$ is suppressed.
We can check such behaviors using the lattice QCD calculation.

\section{Summary and Conclusions}

We have studied the short-range interaction of two-baryon systems with $u$, $d$ and $s$ quarks 
from the viewpoint of $SU(6)_{sf}$ symmetry.
In the color singlet six-quark system, when all the quarks are in the same orbit, the $SU(6)_{sf}$ representation
is restricted to [33] symmetric 490-dim. representation.
By analyzing the [33] components of the two baryon spin-flavor wave functions, one determines the dominant 
state at the short distances.
When the [33] component is a minor contribution in a two-baryon system, then such a system has a strong repulsion
at short distances. This is the case that the state is almost forbidden by the Pauli principle.
Such a almost-Pauli forbidden state is seen for $N\Sigma$ with $(S,I)=(0,1/2)$ and $(1,3/2)$ systems.

In some cases, there are more than one Pauli-allowed [33] states with different flavor representations.
Then for the $BB$ coupled channel system, there are two independent Pauli-allowed states.
In the case of the strangeness $-2$, $S=0$ and $I=0$ system with $\Lambda\Lambda$, $N\Xi$ and so on, 
we find that the Pauli-allowed components of $\Lambda\Lambda$ and $N\Xi$ are orthogonal so that 
the $SU(6)_{sf}$ invariant interaction does not induce the transition. This explains from the symmetry viewpoint that
the $N\Xi$ conversion to $\Lambda\Lambda$ in the $^1S_0$ channel is suppressed. A similar suppression is also predicted in 
the $N\Lambda-N\Sigma$ conversion in the $^3S_1$ channel.

\section*{Acknowledgments}

It is a great honor for me to acknowledge late Prof. Akito Arima for his continuous encouragement during my entire 
career in physics. This paper is based on the research carried out with Akito, Ref.~14, where we formulated the quark 
shell model wave functions for the six-quark ($B=2$) systems. 
This approach happened to be very fruitful and successful in revealing properties of the short-range 
baryon-baryon interactions.

I also acknowledge Drs. Tetsuo Hatsuda, Emiko Hiyama, and Takumi Doi for useful discussions.
This work was supported by Grants-in-Aid for Scientific Research No.~JP20K03959, and No.~JP21H00132.

\appendix
\section{Matrix elements of the spin-flavor exchange operator}

We follow the Nijmegen (de Swart) prescription 
for the $SU(3)_f$ wave functions of the octet and decuplet baryons\cite{dS}.
In order to obtain the $SU(6)_{sf}$ eigenstates, it is convenient to use the matrix elements of the exchange operator,
which interchanges the spin-flavor part of the quark wave functions between the baryons.

For ${\cal S}=-2$, $S=0$ and $I=0$ states,
the matrix elements of the exchange operator $P_{36}^{sf}$ of quarks 3 and 6, assuming that the 1st (2nd) baryon is formed by quarks 1, 2 and 3 (4, 5 and 6).
\begin{eqnarray}
&& \langle P_{36}^{sf}\rangle
=\frac{1}{27}
\begin{pmatrix}
{
0&0&3\sqrt{3} &3\sqrt{6} \cr
0& -3&-4\sqrt{3} & 2\sqrt{6}  \cr
3\sqrt{3}&-4\sqrt{3}& 2&-\sqrt{2}\cr
3\sqrt{6}&2\sqrt{6}& -\sqrt{2} & 1  \cr
}
\end{pmatrix}
\begin{matrix}
{
\Lambda\Lambda \cr N\Xi_S \cr \Sigma\Sigma \cr \Sigma^*\Sigma^* \cr
}
\end{matrix}.
\end{eqnarray}

For ${\cal S}=-1$, $S=0$ and $I=1/2$ states, we have
\begin{eqnarray}
&& \langle P_{36}^{sf}\rangle
=\frac{1}{27}
\begin{pmatrix}
{
0&-3& 6\sqrt{2}\cr
-3&8& 2\sqrt{2}\cr
6\sqrt{2}&2\sqrt{2}& 1\cr
}
\end{pmatrix}
 \begin{matrix}
{
N\Lambda_S \cr N\Sigma_S \cr \Delta\Sigma^*_S \cr
}
\end{matrix}.
\end{eqnarray}

\begin{widetext}
For ${\cal S}=-1$, $S=1$ and $I=1/2$ states, we have
\begin{eqnarray}
&& \langle P_{36}^{sf}\rangle
=\frac{1}{27}
\begin{pmatrix}
{
0&1&2\sqrt{2} &4\sqrt{2} &2\sqrt{10}\cr
1&0& 2\sqrt{2} & 4\sqrt{2} &-2\sqrt{10} \cr
2\sqrt{2}&2\sqrt{2}& 7&-4& 0\cr
4\sqrt{2}&4\sqrt{2}& -4 & 1 &0 \cr
2\sqrt{10}& -2\sqrt{10}&0&0& 1\cr
}
\end{pmatrix}
\begin{matrix}
{
N\Lambda_A \cr N\Sigma_A \cr N\Sigma^*_A \cr \Delta\Sigma_A \cr \Delta\Sigma^*_A\cr
}
\end{matrix}.
\end{eqnarray}

\end{widetext}

For the non-strange $(S,I)=(0,1)$ and $(1,0)$ states, we have
\begin{eqnarray}
&& \langle P_{36}^{sf}\rangle
=\frac{1}{27}
\begin{pmatrix}
{
-1&4\sqrt{5}\cr
4\sqrt{5}&1\cr
}
\end{pmatrix}
 \begin{matrix}
{
NN(I=1) \cr \Delta\Delta(I=1) \cr
}
\end{matrix}\\
&&
 \langle P_{36}^{sf}\rangle
=\frac{1}{27}
\begin{pmatrix}
{
-1&4\sqrt{5}\cr
4\sqrt{5}&1\cr
}
\end{pmatrix}
 \begin{matrix}
{
NN(I=0) \cr \Delta\Delta(I=0) \cr
}
\end{matrix}.
\end{eqnarray}

By diagonalizing the above exchange operator matrix elements, we obtain the irreducible representations of the 
$SU(6)_{sf}$ symmetry.
The eigenvalues are $-1/3$ for the [33] representation and $1/3$ for [51].


\begin{thebibliography}{99}

\bibitem{Oka:1986fr}
M.~Oka, K.~Shimizu and K.~Yazaki,
``Hyperon - Nucleon and Hyperon-hyperon Interaction in a Quark Model,''
Nucl. Phys. A \textbf{464}, 700-716 (1987).
\bibitem{generalized-nuclear-force}
T.~A.~Rijken, M.~M.~Nagels and Y.~Yamamoto,
``Baryon-Baryon Interactions $S = 0, -1, -2, -3, -4$,''
Few Body Syst. \textbf{54}, 801-806 (2013).

\bibitem{hypernuclear physics} 
A.~Gal, E.~V.~Hungerford and D.~J.~Millener,
``Strangeness in nuclear physics,''
Rev. Mod. Phys. \textbf{88}, 035004 (2016).
\bibitem{Hiyama-Nakazawa}
E.~Hiyama and K.~Nakazawa,
``Structure of $S=-2$ Hypernuclei and Hyperon-Hyperon Interactions,''
Ann. Rev. Nucl. Part. Sci. \textbf{68}, 131-159 (2018).

\bibitem{neutronstar} 
W.~Weise,
``Equation of state and strangeness in neutron stars - role of hyperon-nuclear three-body forces -,''
EPJ Web Conf. \textbf{271}, 06003 (2022).
\bibitem{Julich model} 
J.~Haidenbauer, U.~G.~Mei\ss{}ner, 
``Jülich hyperon-nucleon model revisited,''
Phys.\ Rev.\ C \textbf{72}, 044005 (2005).
\bibitem{Nijmegen model} 
T.~A.~Rijken, M.~M.~Nagels and Y.~Yamamoto,
``Baryon-baryon interactions: Nijmegen extended-soft-core models,''
Prog. Theor. Phys. Suppl. \textbf{185}, 14-71 (2010).\\
M.~M.~Nagels, T.~A.~Rijken and Y.~Yamamoto,
``Extended-soft-core baryon-baryon model ESC16. II. Hyperon-nucleon interactions,''
Phys. Rev. C \textbf{99}, 044003 (2019).
\bibitem{chiral EFT} 
S.~Petschauer, J.~Haidenbauer, N.~Kaiser, U.~G.~Mei\ss{}ner and W.~Weise,
``Hyperon-nuclear interactions from SU(3) chiral effective field theory,''
Front. in Phys. \textbf{8}, 12 (2020).
\bibitem{HALQCD} 
N.~Ishii, S.~Aoki and T.~Hatsuda,
``The Nuclear Force from Lattice QCD,''
Phys. Rev. Lett. \textbf{99}, 022001 (2007).\\
T.~Inoue \textit{et al.} [HAL QCD],
``Baryon-Baryon Interactions in the Flavor SU(3) Limit from Full QCD Simulations on the Lattice,''
Prog. Theor. Phys. \textbf{124}, 591-603 (2010).
\bibitem{exotic-HALQCD}
S.~Gongyo, \textit{et al.} [HAL QCD],
``Most Strange Dibaryon from Lattice QCD,''
Phys. Rev. Lett. \textbf{120}, no.21, 212001 (2018).\\
T.~Iritani \textit{et al.} [HAL QCD],
``$N\Omega$ dibaryon from lattice QCD near the physical point,''
Phys. Lett. B \textbf{792}, 284-289 (2019).

\bibitem{YNexp} 
K.~Miwa, \textit{et al.}
``Recent progress and future prospects of hyperon nucleon scattering experiment,''
EPJ Web Conf. \textbf{271}, 04001 (2022).
\bibitem{Femtoscopy} 
L.~Fabbietti, V.~Mantovani Sarti and O.~Vazquez Doce,
``Study of the Strong Interaction Among Hadrons with Correlations at the LHC,''
Ann. Rev. Nucl. Part. Sci. \textbf{71}, 377-402 (2021).


\bibitem{QCM} 
M.~Oka and K.~Yazaki,
``Nuclear Force in a Quark Model,''
Phys. Lett. B \textbf{90}, 41-44 (1980).\\
M.~Oka, K.~Shimizu and K.~Yazaki,
``Quark cluster model of baryon baryon interaction,''
Prog. Theor. Phys. Suppl. \textbf{137}, 1-20 (2000).
\bibitem{Arima}
S.~Ohta, A.~Arima, K.~Yazaki and M.~Oka,
``A shell model study of six quark system,''
Phys. Lett. B \textbf{119}, 35-38 (1982)
[erratum: Phys. Lett. B \textbf{123}, 477-477 (1983)].
\bibitem{Harvey}
M.~Harvey,
``Effective nuclear forces in the quark model with Delta and hidden color channel coupling,''
Nucl. Phys. A \textbf{352}, 326-342 (1981).
\bibitem{Faessler} 
A.~Faessler, F.~Fernandez, G.~Lubeck and K.~Shimizu,
``The Quark Model and the Nature of the Repulsive Core of the Nucleon Nucleon Interaction,''
Phys. Lett. B \textbf{112}, 201-205 (1982).

\bibitem{Jaffe} 
R.~L.~Jaffe,
``Perhaps a Stable Dihyperon,''
Phys. Rev. Lett. \textbf{38}, 195-198 (1977)
[erratum: Phys. Rev. Lett. \textbf{38}, 617 (1977)].
\bibitem{H-search} 
J.~K.~Ahn [J-PARC E42],
``Search for the H-dibaryon near $\Lambda\Lambda$ and $\Xi^-p$ thresholds at J-PARC,''
PoS \textbf{Hadron2017}, 124 (2018).
\bibitem{H-dibaryon-HALQCD}
T.~Inoue \textit{et al.} [HAL QCD],
``Bound H-dibaryon in Flavor SU(3) Limit of Lattice QCD,''
Phys. Rev. Lett. \textbf{106}, 162002 (2011).

\bibitem{OSYforH} 
M.~Oka, K.~Shimizu and K.~Yazaki,
``The Dihyperon State in the Quark Cluster Model,''
Phys. Lett. B \textbf{130}, 365-368 (1983).\\
S.~Takeuchi and M.~Oka,
``Can the H particle survive instantons?,''
Phys. Rev. Lett. \textbf{66}, 1271-1274 (1991).\\
T.~Sakai, K.~Shimizu and K.~Yazaki,
``H dibaryon,''
Prog. Theor. Phys. Suppl. \textbf{137}, 121-145 (2000).
\bibitem{Oka-flavor-octet}	
M.~Oka,
``Flavor Octet Dibaryons in the Quark Model,''
Phys. Rev. D \textbf{38}, 298 (1988).

\bibitem{XiN-LL conversion} 
K.~Sasaki \textit{et al.} [HAL QCD],
``$\Lambda\Lambda$ and N$\Xi$ interactions from lattice QCD near the physical point,''
Nucl. Phys. A \textbf{998}, 121737 (2020).\\
E.~Hiyama, K.~Sasaki, T.~Miyamoto, T.~Doi, T.~Hatsuda, Y.~Yamamoto and T.~A.~Rijken,
``Possible lightest $\Xi$ Hypernucleus with Modern $\Xi N$ Interactions,''
Phys. Rev. Lett. \textbf{124}, no.9, 092501 (2020).

\bibitem{Xi-exp} 
K.~Nakazawa, \textit{et al.}
``The first evidence of a deeply bound state of $\Xi-{}^{14}$N system,''
PTEP \textbf{2015}, no.3, 033D02 (2015).\\
S.~H.~Hayakawa \textit{et al.} [J-PARC E07],
``Observation of Coulomb-Assisted Nuclear Bound State of $\Xi-{}^{14}$N System,''
Phys. Rev. Lett. \textbf{126}, 062501 (2021).



\bibitem{Inoue:2018axd}
T.~Inoue [HAL QCD],
``Strange Nuclear Physics from QCD on Lattice,''
AIP Conf. Proc. \textbf{2130}, 020002 (2019).
\bibitem{Nemura:2018tay}
H.~Nemura [HALQCD],
``Hyperon-nucleon interaction from lattice QCD at ${(m_\pi,m_K)\approx(146,525)}$ MeV,''
AIP Conf. Proc. \textbf{2130}, 040005 (2019).

\bibitem{J-PARCE40:2022nvq}
T.~Nanamura \textit{et al.} [J-PARC E40],
``Measurement of differential cross sections for $\Sigma+p$ elastic scattering in the momentum range 0.44\textendash{}0.80\,GeV/c,''
PTEP \textbf{2022}, no.9, 093D01 (2022)


\bibitem{Sigma-Nucleus} 
E.~Friedman and A.~Gal,
``In-medium nuclear interactions of low-energy hadrons,''
Phys. Rept. \textbf{452}, 89-153 (2007).

\bibitem{DGG}
A.~De Rujula, H.~Georgi and S.~L.~Glashow,
``Hadron Masses in a Gauge Theory,''
Phys. Rev. D \textbf{12}, 147-162 (1975).

\bibitem{OT}
M.~Oka and S.~Takeuchi,
``Instanton Induced Quark Quark Interactions in Two Baryon Systems,''
Phys. Rev. Lett. \textbf{63}, 1780-1783 (1989),\\
E.~V.~Shuryak and J.~L.~Rosner,
``Instantons and Baryon Mass Splittings,''
Phys. Lett. B \textbf{218}, 72-74 (1989).


\bibitem{dS} 
J.~J.~de Swart,
``The Octet model and its Clebsch-Gordan coefficients,''
Rev. Mod. Phys. \textbf{35}, 916-939 (1963)
[erratum: Rev. Mod. Phys. \textbf{37}, 326-326 (1965)].

\end{thebibliography}
\end  {document}